\documentstyle[12pt]{article}

\newcommand{\be}{\begin{equation}}
\newcommand{\ee}{\end{equation}}
\newcommand{\Dlt}{\Delta}
\newcommand{\dlt}{\delta}

\newcommand{\br}{{\bf r}}

\newcommand{\bt}{\beta}
\newcommand{\vp}{\varphi}
\newcommand{\ep}{\varepsilon}
\newcommand{\al}{\alpha}
\newcommand{\ra}{\rightarrow}

\newcommand{\gm}{\gamma}
\newcommand{\om}{\omega}

\begin{document}

\begin{center}   

{\Large{\bf Generation of nonground-state condensates and 
adiabatic paradox} \\ [5mm]

V.I. Yukalov$^{1,*}$ and V.S. Bagnato$^2$} \\ [3mm]

{\it
$^1$Bogolubov Laboratory of Theoretical Physics, \\
 Joint Institute for Nuclear Research, Dubna 141980, Russia \\ [3mm]

$^2$Instituto de Fisica de S\~ao Carlos, Universidade de S\~ao Paulo, \\
 Caixa Postal 369, 13560-970 S\~ao Carlos, SP, Brazil}

\end{center}

\vskip 1cm

\begin{abstract}

The problem of resonant generation of nonground-state 
condensates is addressed aiming at resolving the seeming
paradox that arises when one resorts to the adiabatic 
representation. In this picture, the eigenvalues and 
eigenfunctions of a time-dependent Gross-Pitaevskii 
Hamiltonian are also functions of time. Since the level  
energies vary in time, no definite transition frequency 
can be introduced. Hence no external modulation with a 
fixed frequency can be made resonant. Thus, the resonant 
generation of adiabatic coherent modes is impossible. 
However, this paradox occurs only in the frame of the 
adiabatic picture. It is shown that no paradox exists in 
the properly formulated diabatic representation. The 
resonant generation of {\it diabatic} coherent modes is
a well defined phenomenon. As an example, the equations 
are derived, describing the generation of diabatic coherent 
modes by the combined resonant modulation of the trapping 
potential and atomic scattering length.

\end{abstract}

\vskip 2cm


\vskip 2cm


\vskip 3cm 
$^*$Corresponding author: e-mail: yukalov@theor.jinr.ru

\newpage

\section{Origin of adiabatic paradox}

Nonequilibrium ultracold Bose-condensed gases are well described 
by the time-dependent Gross-Pitaevskii equation [1-10], which is 
a nonlinear Schrodinger equation
\be
\label{1}
i\dot{\psi}(t) = H[\psi,t] \; \psi(t) \; .
\ee
Here $\hbar\equiv 1$ and, to simplify the notation, the condensate 
wave function $\psi(t)$ is assumed to be a vector in spatial variables 
and the nonlinear Hamiltonian $H[\psi,t]$ is a matrix in these variables.
The overdot means the differentiation with respect to time $t$. The 
temperature is close to zero, so that all $N$ atoms are supposed to be 
in the coherent Bose-condensed state. Generally, the condensate wave 
function is normalized to the total number of condensed atoms. If this 
function is denoted as $\eta(t)$, it is always possible to introduce, 
by means of the relation
$$
\eta(t) = \sqrt{N} \; \psi(t) \; ,
$$
the function $\psi(t)$ that is normalized to one,
\be
\label{2}
|| \psi(t) ||^2 \; \equiv \; < \psi(t)|\psi(t) > \; = 
\; 1\; .
\ee
It is the latter function that is assumed in Eq. (1). The system 
of trapped atoms is subject to the action of an externally  
induced modulation field, so that the nonlinear Hamiltonian
$H[\psi,t]$ depends on time directly as well as through the 
function $\psi(t)$. 

It is possible to consider the adiabatic eigenvalue problem
\be
\label{3}
H[\psi_n,t] \; \psi_n(t) = E_n(t) \psi_n(t) \; ,
\ee
treating time as a fixed parameter. Then, of course, the 
eigenfunctions $\psi_n(t)$ and the eigenvalues $E_n(t)$ depend 
on time. The multi-index $n$ labelles the eigenfunctions and 
eigenvalues. The energy spectrum, for trapped atoms is discrete.
The eigenfunctions can be normalized to one:
\be
\label{4}
|| \psi_n(t) ||^2 \; \equiv \; < \psi_n(t)|\psi_n(t) > \; = 
\; 1\; .
\ee
These functions correspond to the {\it adiabatic coherent modes}.

The nonlinear Eq. (1) can possess different solutions. A wide and 
important class of solutions can be represented as the expansion 
over the adiabatic coherent modes:
\be
\label{5}
\psi(t) = \sum_n a_n(t) \; \exp\{ i\chi_n(t) \} \;
\psi_n(t) \; ,
\ee
where the phase 
\be
\label{6}
\chi_n(t) = \dlt_n(t) + \zeta_n(t)
\ee
is the sum of the dynamic phase
\be
\label{7}
\dlt_n(t)  \equiv - \int_0^t E_n(t') \; dt'
\ee
and of the geometric phase
\be
\label{8}
\zeta_n(t) \equiv i \int_0^t < \psi_n(t') | \dot{\psi}_n(t') >
dt' \; .
\ee
Formula (5) does not exhaust all admissible solutions of the 
nonlinear Schrodinger equation (1). But here we are interested in 
the class of solutions that are representable as expansion (5). 
   
The problem of dealing with the nonlinear Hamiltonian $H[\psi,t]$ 
is that the form $H[\psi_n,t]$ is not a Hermitian operator since
\be
\label{9}
< \psi_m(t)\; | \; H[\psi_n,t]\psi_n(t) > \; \neq \; 
< H[\psi_n,t]\psi_m(t)\; |\; \psi_n(t) > \; .
\ee
Therefore, the eigenfunctions $\psi_m(t)$ and $\psi_n(t)$ for 
$m \neq n$, generally, are not orthogonal,
$$
< \psi_m(t) | \psi_n(t) > \; \neq \; \dlt_{mn} \; .
$$
These functions do not necessarily compose a complete basis. 
However, in the space that is a closed linear envelope  
$Span \{\psi_n(t)\}$, they form a total normalized basis, so that 
the functions from this space can be represented in the form of 
expansion (5). Substituting the latter into Eq. (1) yields
$$
\sum_m \left [ \dot{a}_m(t) < \psi_n(t)\; |\; \psi_m(t) > \; + \;
a_m(t) < \psi_n(t)\; |\; \dot{\psi}_m(t) > (1 -\dlt_{mn}) \; + \right.
$$
\be
\label{10}
\left.
+ \; i a_m(t) < \psi_n(t)\; |\;  H[\psi,t] - E_m(t)\; |\; \psi_m(t) > 
\right ] \; \exp\{ i \chi_m(t) \} \; = \; 0 \; .
\ee
This equation is difficult to simplify because of the 
nonorthogonality of the basis $\{\psi_n(t)\}$.

If the time dependence in the Hamiltonian $H[\psi,t]$ enters 
through an external alternating field with a fixed frequency 
$\omega$, then the latter cannot be tuned to resonance with any 
of the time-dependent transition frequencies $E_m(t) - E_n(t)$.
Moreover, the eigenenergies enter expansion (5) not explicitly 
but through the dynamic phases (7). Thus, no resonance 
condition can be defined, making the resonant generation of 
adiabatic coherent modes impossible. It is this argument that 
one raises against the possibility of generating nonground-state 
condensates of trapped atoms.

\section{Paradox-free diabatic representation}

Now we show that no paradox arises in a properly formulated 
diabatic representation. Let the Hamiltonian be a sum
\be
\label{11}
H[\psi,t] = H_0 [\psi] + V[\psi,t] \; ,
\ee
where the first term describes the system of cold trapped atoms,  
while the second term contains the direct time dependence from 
externally induced modulation fields.

The {\it diabatic coherent modes} are defined as the solutions 
to the {\it time-independent} eigenproblem
\be
\label{12}
H_0[ \vp_n] \; \vp_n \; = \; E_n \; \vp_n \; .
\ee
The definition of the diabatic coherent modes as time-independent, 
{\it stationary} solutions makes them principally different from 
the adiabatic modes of Sec. 1. It is this definition of coherent 
modes that was introduced in Refs. [11-13]. Particular examples 
of such modes are vortices [14], though many other types of modes
are admissible [11-13]. The lowest diabatic coherent mode 
corresponds to the usual ground-state Bose-Einstein condensate. 
While the excited diabatic coherent modes represent 
nonground-state condensates.

The nonlinear Hamiltonian $H_0[\varphi_n]$ is not Hermitian. Hence 
the eigenmodes $\varphi_m$ and $\varphi_n$, with $m \neq n$, are not 
orthogonal, although all of them can be normalized to one:
\be
\label{13}
|| \vp_n ||^2 \; \equiv \; < \vp_n | \vp_n > \; = \; 1 \; .
\ee
But again, in the space that is a closed linear envelope 
$Span\{\varphi_n\}$, the family $\{\varphi_n\}$ forms a normalized 
total basis. In this space, the corresponding class of 
solutions to Eq. (1) can be represented as the expansion
\be
\label{14}
\psi(t)  =\sum_n \; c_n(t) \exp ( - i E_n t) \vp_n
\ee
over the stationary coherent modes.

Substituting expansion (14) into Eq. (1) results in the equation
$$
i \sum_m \; \dot{c}_m(t) < \vp_n| \vp_m>
\exp ( - i \om_{mn}t ) = 
$$
\be
\label{15}
= \sum_m \; c_m(t) < \vp_n \; | \;
H[\psi,t] - E_n \; |\; \vp_m > \exp( - i \om_{mn} t ) \; ,
\ee
in which
\be
\label{16}
\om_{mn} \equiv E_m - E_n \; .
\ee
The latter expression defines transition frequencies that do no 
depend on time. Consequently, a given frequency $\omega$ of an  
external alternating field can be tuned to the resonance with 
one of these transition frequencies.

Equation (15) can be simplified assuming that the coefficients
$c_n(t)$ are slow functions of time as compared to the fastly 
oscillating exponentials, so that
\be
\label{17}
\left | \frac{\dot{c}_n(t)}{E_n} \right | \; \ll \; 1 \; .
\ee
Then, we can use the averaging method [15] and the scale 
separation approach [16-18]. Employing these techniques, we 
average Eq. (15) over time, defining the time averaging as
\be
\label{18}
\{ f(t) \}_{av} \equiv \lim_{\tau\ra\infty} \;
\frac{1}{\tau} \; \int_0^\tau f(t)\; dt \; ,
\ee
and treating the coefficients $c_n(t)$ as quasi-integrals of 
motion. The averaging of the exponentials gives
$$
\{ \exp ( - i \om_{mn} t) \}_{av}  =\dlt_{mn} \; .
$$ 
Introducing the notation
\be
\label{19}
\kappa_{nm}(t) \equiv \left \{ \; < \vp_n \; | \; H [\psi,t] -
E_m \; |\; \vp_m> \exp ( -i\om_{mn} t) \; \right \}_{av} \; ,
\ee
we come to the equation
\be
\label{20}
i \dot{c}_n(t)  = \sum_m \; \kappa_{nm}(t) \; c_m(t) \; .
\ee
The solutions $c_n(t)$ define the time variation of the 
fractional mode populations $|c_n(t)|^2$.

The principal difference of the diabatic representation, 
described in this section, from the adiabatic picture of  
Sec. 1, is as follows: 

\begin{itemize}
\item
There exist well defined stationary energy levels of the 
diabatic coherent modes.
\item
The frequency $\omega$ of an external field can be tuned to 
the resonance with one of the transition frequencies $\om_{mn}$.
\item
The diabatic coherent modes, being the solutions to the 
stationary eigenproblem, do not depend on time.
\item
The equation for the coefficient functions of expansion (14)
can be simplified by means of the averaging techniques.   
\end{itemize}

These essential points make it possible to consider well 
defined resonance conditions. Consequently, it is feasible 
to realize the resonance generation of the {\it diabatic 
coherent modes}.

\section{Generation of coherent modes}

To specify the above equations, let us consider the atoms 
interacting through the local potential
\be
\label{21}
\Phi(\br) = \Phi_0 \dlt(\br) \qquad
\left ( \Phi_0  \equiv 4\pi \; \frac{a_s}{m} 
\right ) \; ,
\ee
where $m$ is atomic mass and $a_s$, scattering length.
The nonlinear Hamiltonian takes the form
\be
\label{22}
H [\psi, t ] = -\; \frac{\nabla^2}{2m} + U(\br,t) + 
N\Phi(t) | \psi(\br,t)|^2 \; ,
\ee
where the term
\be
\label{23}
U(\br,t) = U(\br) + V(\br,t)
\ee
consists of a trapping potential $U(r)$ and of an external 
driving field $V(r,t)$. The interaction part
\be
\label{24} 
\Phi(t) = \Phi_0 + \ep(t) 
\ee
contains the static interaction potential $\Phi_0$, defined 
in Eq. (21), and an additional term describing a possible 
modulation of the scattering length by means of the Feshbach 
resonance technique [19].

The generation of coherent modes by modulating the trapping
potential, as in Eq. (23), their properties, and different 
applications were considered in Refs. [11-13,20-35]. The 
analytical treatment, using the averaging method, was 
compared and found to be in good agreement with the direct 
simulation of the Gross-Pitaevskii equation [36,37]. The 
possibility of the resonant creation of coherent modes by 
modulating the atomic scattering length using Feshbach 
resonance was considered in Ref. [38]. This method can be 
preferred for those atomic species that demonstrate high 
tunability of their scattering length [39]. In the present 
paper, we take into account both these ways, simultaneously  
modulating the trapping potential as well as the scattering 
length. By combining both these techniques can give 
additional advantage for generating nonground-state 
condensates. For instance, the combination of these two 
techniques can be employed for generating two different excited  
coherent modes, in addition to the ground-state one, or for 
enhancing the generation of the chosen mode.       

Hamiltonian (22) can be presented in form (11), with
\be
\label{25}
H_0[\psi]  = -\; \frac{\nabla^2}{2m} + U(\br) + 
N\Phi_0 | \psi(\br,t) |^2
\ee
and 
\be
\label{26}
V[\psi,t] = V(\br,t) + N\ep(t) |\psi(\br,t) |^2 \; .
\ee
Eigenproblem (12) becomes
\be
\label{27}
H_0 [\vp_n] \; \vp_n(\br) = E_n \; \vp_n(\br) \; .
\ee
And expansion (14) takes the form
\be
\label{28}
\psi(\br,t) = \sum_n \; c_n(t) e^{-iE_nt} \vp_n(\br) \; .
\ee
Then we should follow the consideration of Sec. 2.

To explicitly accomplish the averaging procedure of Eq. (18), 
we have to concretize the trap modulation and the type of the 
interaction alternation. We take the general form of the 
oscillating trapping modulation
\be
\label{29}
V(\br,t) =  V_1(\br)\cos \om t + V_2(\br) \sin \om t
\ee
and the similar type of the interaction alternation
\be
\label{30}
\ep(t) =\ep_1 \cos \om t + \ep_2 \sin \om t \; .
\ee
The oscillation frequency is taken the same for both 
modulations, so that the same nonground-state coherent mode be 
excited. In principle, two different frequencies could be taken 
for modulations (29) and (30), if we would wish to excite two 
different nonground-state modes simultaneously. The frequency 
$\omega$ is tuned to the resonance with a chosen transition 
frequency 
\be
\label{31}
\om_{21} \equiv E_2 - E_1 \; .
\ee
If the generation procedure starts with an equilibrium 
Bose-Einstein condensate, then $E_1$ is the lowest energy of 
atoms in the trap, corresponding to the condensate chemical 
potential [10]. The resonance condition reads as
\be
\label{32}
\left | \frac{\Dlt\om}{\om_{21} } \right | \; \ll \; 1 \qquad
( \Dlt\om \equiv \om -\om_{21} ) \; .
\ee

For what follows, it is convenient to introduce the notation for 
the transition amplitudes 
$$
\al_{mn} \equiv N \Phi_0 \int | \vp_m(\br)|^2 \left [ 2 
| \vp_n(\br)|^2 - | \vp_m(\br)|^2 \right ] \; d\br \; ,
$$
$$
\bt_{mn} \equiv \int \vp_m^*(\br)\; [ V_1(\br) - 
i V_2(\br) ] \; \vp_n(\br) \; d\br \; ,
$$
\be
\label{33}
\gm_n \equiv N (\ep_1 - i\ep_2) \; \int \vp_1^*(\br) \;
|\vp_n(\br)|^2 \vp_2(\br) \; d\br \; .
\ee
The first of them is the amplitude due to atomic interactions, 
the second one is caused by the trap modulation, and the third 
amplitude is due to the interaction oscillation. To satisfy 
condition (17), these amplitudes are to be small, such that
$$
\left | \frac{\al_{mn}}{\om_{mn} } \right | \; \ll \; 1 \; , \qquad
\left | \frac{\bt_{mn}}{\om_{mn} } \right | \; \ll \; 1 \; , \qquad
\left | \frac{\gm_n}{\om_{mn} } \right | \; \ll \; 1 \; , 
$$
where $\omega_{mn}\neq 0$.

Substituting expressions (29) and (30) into Eq. (20), under the 
resonance condition (32), yields the evolution equations for the 
coefficient functions $c_n = c_n(t)$ defining the fractional mode 
populations
\be
\label{34}
p_n(t) \equiv | c_n(t)|^2 \; .
\ee
Under the resonance condition (32), the system of equations 
reduces to only two equations for the considered coherent modes:
$$
i \; \frac{dc_1}{dt} = \al_{12} | c_2|^2 c_1 + 
\frac{1}{2} \left (2\gm_1 | c_1|^2 + \gm_2 |c_2|^2 + 
\bt_{12} \right ) c_2 e^{i\Dlt\om t} +
\frac{1}{2} \; \gm_1^* c_2^* c_1^2\; e^{-i\Dlt\om t} \; ,
$$
\be
\label{35}
i \; \frac{dc_2}{dt} = \al_{21} | c_1|^2 c_2 + 
\frac{1}{2} \left ( 2\gm_2^* | c_2|^2 + \gm_1^* |c_1|^2 + 
\bt_{12}^* \right ) c_1 e^{-i\Dlt\om t} +
\frac{1}{2} \; \gm_2 c_1^* c_2^2 \; e^{i\Dlt\om t} \; .
\ee
Solving these equations gives the dynamics of the fractional mode 
populations (34).

Here we have considered the case of zero temperature and 
asymptotically weak atomic interactions, when the whole atomic 
cloud is in a coherent state described by the Gross-Pitaevskii 
equation. It is possible to extend the consideration to the case 
of finite temperatures and interactions, when, in addition to the
fraction of condensed atoms, there exists a fraction of 
uncondensed atoms. This can be done by invoking, for instance, a  
stochastic variant of the Gross-Pitaevskii equation [40-42] or by 
employing the fully self-consistent theory [43-45]. The 
possibility of generating nonground-state condensates, even at 
finite temperatures and interactions, is due to the resonant 
nature of the generation procedure; since the energy levels of 
the coherent modes in a trap are discrete, while the spectrum of 
uncondensed atoms is continuous [43].     

Equations (35) describe the dynamics of the guiding centers 
that correspond to the first approximation of the averaging 
techniques [15--18]. In this approximation, under the resonance 
condition (32), the system of equations for the functions $c_n(t)$ 
reduces to only two Eqs. (35), if at the initial time $t=0$ other 
modes were not populated, so that
$$
c_n(0) = 0 \qquad (n\neq 1,2) \; .
$$
Then $c_n(t)=0$, if $n\neq 1,2$, for all $t>0$. Generally, there 
also exist nonresonant transitions between the coherent modes. 
These nonresonant transitions can be taken into account by the 
higher-order approximations of the scale separation approach 
[16--18]. In the higher-order approximations, the mode amplitudes 
$c_n(t)$ for the nonresonant modes, for which $n\neq 1,2$, 
become nonzero, even if at the initial time these modes were 
not populated. However, the amplitudes of these nonresonant 
modes remain small during the time interval $0\leq t<t_{res}$, 
when they can be neglected. But for longer times $t>t_{res}$, 
the nonresonant modes cannot be neglected. Thus, the resonant 
generation of coherent modes is limited by the time $t_{res}$, 
which was estimated in Refs. [25,43]. By the order of magnitude, 
the resonance time $t_{res}$ can be made comparable to the 
lifetime of atoms in a trap, provided the absolute values of 
the transition amplitudes $|\al_{12}|$, $|\bt_{12}|$, $|\gm_{1}|$, 
and $|\gm_{2}|$ are much smaller than the transition frequency 
$\om_{21}\approx\om$. Thus, Eqs. (35) are valid only for $t<t_{res}$. 
But the time $t_{res}$, being of the order of the atom lifetime 
in a trap, is sufficiently long for realizing the process of 
the resonant mode generation.

In conclusion, we have demonstrated that the appearance of the 
adiabatic paradox, precluding the generation of the adiabatic 
modes, exists solely in the adiabatic picture. The adiabatic 
representation is unsuitable for the case of resonance. However, 
the diabatic representation, developed in Sec. 2, contains no 
paradoxes and is perfectly appropriate for describing the 
resonant generation of coherent modes characterizing 
nonground-state condensates. Such a generation can be realized 
by the resonant modulation of either the trapping potential, or 
the scattering length, or by the combined modulation of both.    

The appearance of the resonantly generated nonground-state 
condensates can be observed, for instance, by means of collective 
light scattering [46--49]. Another way of their registration is 
through the time-of-flight experiments, as is discussed in Ref. [50].
    
\vskip 5mm

{\it Acknowledgements}. Useful discussions with E.P. Yukalova 
are appreciated. Financial support from the Russian Foundation 
for Basic Research (Grant 08-02-00118) is acknowledged.

\end{document}